# Comment: Bayesian Checking of the Second Levels of Hierarchical Models

**Valen E. Johnson**

This article extends Bayarri and Berger's (1999) proposal for model evaluation using "partial posterior" $p$ values to the evaluation of second-stage model assumptions in hierarchical models. Applications focus on normal-normal hierarchical models, although the final example involves an application to a beta-binomial model in which the distribution of the test statistic is assumed to be approximately normal.

The notion of using partial posterior $p$ values is potentially appealing because it avoids what the authors refer to as "double use" of the data, that is, use of the data for both fitting model parameters and evaluating model fit. In classical terms, this phenomenon is synonymous to masking and is widely known to reduce the power of test statistics for diagnosing model inadequacy. In the present context, masking is avoided by defining the reference distribution of a test statistic $t$ by the partial posterior distribution, defined as

$$(1) \qquad \pi(\theta \mid x_{obs}/t_{obs}) \propto \frac{f(x_{obs} \mid \theta)\pi(\theta)}{f(t_{obs} \mid \theta)}.$$

Heuristically, the partial posterior distribution contains information in the data $x_{obs}$ about model parameter $\theta$ not reflected in $t_{obs}$. From this definition, it follows that the partial posterior distribution and (full) posterior distribution are equivalent when $t$ is ancillary, and that the partial posterior distribution and prior distribution coincide when $t$ is sufficient. The latter fact suggests that partial posterior distributions defined with respect to improper prior densities may not be proper when the test statistic is "approximately sufficient" for some subset of parameter values. It also precludes the use of partial posterior model assessment for objective Bayesian models using test statistics that are sufficient, although the authors presumably regard sufficient test statistics as being useful only for assessing the adequacy of (proper) prior distributions. Nonetheless, insight regarding the relative advantages of the proposed methodology as test statistics vary from being "nearly sufficient" to "nearly ancillary" would be useful.

*Valen E. Johnson is Professor of Biostatistics, University of Texas MD Anderson Cancer Center, 515 Holcombe Blvd, Unit 447 Houston, Texas 77030-4009, USA e-mail: ejohnson@mdanderson.org.*

Under regularity assumptions specified in Robins, van der Vaart and Ventura (2000), partial posterior $p$ values also have the important property of being asymptotically uniformly distributed under the null model. Prior-predictive $p$ values and their extensions to $p$ values based on pivotal quantities (described below) share this property—even in finite samples. $p$ values based on posterior predictive and related reference distributions do not, which makes it difficult to interpret these diagnostics for purposes of formal model assessment. Bayarri and Costellanos (B&C) provide convincing examples that illustrate this difficulty and highlight the dangers associated with the naive use of nonuniform $p$ values. However, it should be noted that the extreme $p$ values reported by the authors are perhaps also somewhat suspect given the relatively small sample sizes considered in the examples. That is, even ignoring errors associated with the numerical approximation of the partial posterior density and the resulting distribution of the test statistic, asymptotic uniformity of the partial posterior $p$ values may not have been achieved to the level of accuracy required for the report of partial posterior $p$ values down to the number of significant digits provided. This concern is heightened by the plots in the third column of Figure 1, which suggest that partial posterior $p$ values are anticonservative for moderate sample sizes.

The significant advantage of partial posterior $p$ values—that of reducing masking—does not come without cost, and two potentially difficult tasks must be performed to construct these diagnostics. First, it is necessary to estimate the sampling density of the chosen test statistic as a function of the model parameter $\theta$. In the article, this task is performed only for cases in which the sampling density of the test statistic can be easily approximated by exploiting a translation-invariance property of the normal distribution. Such a strategy is unlikely to work outside of normal family problems or for more sophisticated test statistics (e.g., the $\chi^2$ discrepancy function advocated in Gelman, Meng and Stern, 1996, or the Shapiro–Wilks test statistic illustrated below).

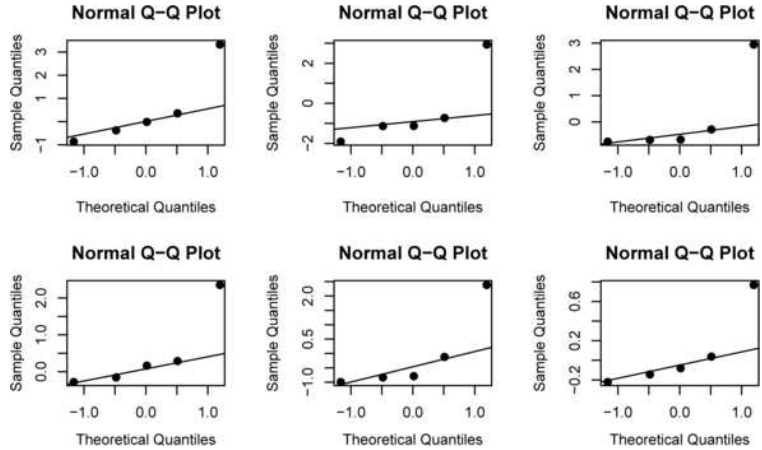

Fig. 1. *Quantile-quantile plots of group mean residuals. The top row depicts the qq-plot obtained from three posterior draws from the model utilizing O'Hagan's prior, while the bottom row depicts qq-plots derived from the three draws from the posterior defined using the truncated version of B&C's improper prior specification.*

Second, the partial posterior distribution function of the test statistic must be evaluated at its observed value. Because the partial posterior distribution is proportional to the ratio of the posterior distribution based on the full data to the sampling distribution of the test statistic determined in the previous step, performing numerical simulations to obtain the value of the partial posterior distribution function at the observed test statistic is also likely to be troublesome. Indeed, even for what are very simple hierarchical models, the authors felt obliged to provide appendices describing the MCMC algorithms used to perform these calculations.

Partial posterior methods also do not seem well-suited for the construction of diagnostic plots. Graphical diagnostics—which are critical to model criticism and exploration—often involve the display of transformations of all data values, and thus are functions of a sufficient statistic. As noted above, this makes the use of partial posterior methods inappropriate for the construction of such plots and may limit the utility of this approach in the exploratory stages of model refinement.

A final point that should be considered in the application of partial posterior $p$ values involves the trade-off between the cost of computing these diagnostics versus the cost of fitting expanded models that have been targeted to detect a particular deviation from the null model. The example in Section 4 illustrates this point well. In that example, a normal-normal hierarchical model with a fixed second-stage mean $\mu_0$ is assumed. By conditioning on a test statistic that represents a component of the sufficient statistic that would be used to estimate $\mu_0$ (if its value was not known a priori), partial posterior model diagnostics overcome the masking effect that plagues the other methods considered in the article. However, fitting an expanded model in which $\mu_0$ was regarded as random would be several orders of magnitude easier to implement. It would also provide a much cleaner summary of the original model's inadequacy. Although this stylized example was only proposed for purposes of illustration, I suspect that similar comments might also apply to more elaborate models.

As it happens, many of the obstacles associated with implementing partial posterior model diagnostics can be overcome by instead defining model diagnostics using pivotal quantities. Like partial posterior model diagnostics, Bayesian model diagnostics based on pivotal quantities also produce test statistics that have a known reference distribution. The primary drawback of diagnostics based on pivotal quantities is that the joint distribution of piv-otal quantities drawn from the same posterior distribution must be evaluated using prior-predictive methodology. However, in many cases the reliance on prior-predictive assessment can be circumvented through the use of probabilistic bounds on distributions of dependent order statistics.

The advantages of diagnostics based on pivotal quantities stem from the fact that the distribution of a pivotal quantity, say $S(x,\theta)$, is the same whether it is evaluated at the "true" (i.e., data-generating) value of the parameter or at a value of $\theta$ drawn from the posterior distribution (Johnson, 2007). Furthermore, many pivotal test statistics are insensitive to the choice of end-stage prior distributions in hierarchical models, which makes their use for diagnostics in such settings straightforward. To illustrate these diagnostics and to demonstrate how they can be used to complement information contained in partial posterior $p$ values, two of the examples considered in B&C are re-evaluated below using diagnostics based on pivotal quantities.

The first example concerns the data and model taken from O'Hagan (2003). From the normal-normal hierarchical structure of this model, it follows that the components of the pivotal vectors

(2) $\quad \epsilon_j = \left\{ \dfrac{y_{ij} - \theta_i}{\sigma_i} \right\} \quad \text{and} \quad E = \left\{ \dfrac{\theta_i - \mu}{\tau} \right\}$

are marginally distributed as independent, standard normal deviates when evaluated at parameter values drawn from the posterior distribution, provided only that proper prior distributions are assumed for the hyperparameters $(\mu, \tau)$.

Two end-stage priors were assumed for the hyperparameters $(\mu, \tau)$ in B&G, one an improper prior and the second the informative prior proposed by O'Hagan. To replicate findings for the improper priors, I assume a priori that

$$\mu \sim U(-a, a), \quad \pi(\sigma^2) \propto \dfrac{1}{\sigma^2} \mathcal{I}(1/a, a)$$

and

$$\pi(\tau^2) \propto \dfrac{1}{\tau} \mathcal{I}(1/a, a),$$

independently for a sufficiently large value of $a$. Although the effect of the value of $a$ (or, more generally, the limiting process used to obtain an improper prior specification) on prior-predictive assessment of the *joint* distribution of pivotal quantities is a topic of active research, the marginal distribution of pivotal quantities obtained for a fixed data vector is generally insensitive to this choice.

Quantile-quantile plots of three values of $E$ for posterior draws of $(\mu, \tau)$ under the proper and improper prior specifications appear in Figure 1. A visual examination of these plots clearly suggests that the fifth group mean is problematic. In practice, the evidence provided by these plots—which are typical of plots obtained for arbitrary draws of $(\mu, \tau)$ from the posterior—would be sufficient to trigger an examination of the distribution of observations from the fifth group.

The notion of formal Bayesian model assessment using $p$ values is a bit oxymoronic, but in the event that a Bayesian $p$ value is desired to more formally assess the adequacy of these models, samples of pivotal vectors like those displayed in Figure 1 can also be used to construct a summary test statistic. For normal data, the Shapiro–Wilks statistic (1965) is an attractive choice for this purpose.

Figure 2 displays histogram estimates of the posterior distribution on the $p$ values obtained by applying the Shapiro–Wilks test statistic to a sample of 50,000 pivotal vectors $E$ obtained from the posterior distributions defined using both the proper and improper prior specifications. Note that under the assumed model assumptions, the marginal distribution of each of the $p$ values displayed in this figure are (exactly) uniform.

In general, prior predictive methods are required to formally evaluate the joint distributions of pivotal quantities like those displayed in the plots of Figure 2. However, prior predictive methods are relatively computationally expensive to implement. As B&C note, they also do not apply to models defined using improper prior distributions. To avoid such computations, bounds on order statistics from dependent variables (Caraux and Gascuel, 1992; Rychlik, 1992) can instead be used to obtain a bound on the $p$ value associated with the joint distribution of a pivotal quantity. In this case, such bounds can be used to obtain a $p$ value for the test of the null hypothesis that the $p$ values obtained from the Shapiro–Wilk statistic were generated from the assumed model (Johnson, 2007). For the proper and (limiting) improper prior specifications, these bounds are $p < 0.07$ and $p < 0.05$, respectively. Note that both of these bounds, as well as the diagnostic plots provided in Figure 1, were obtained using only posterior samples from the assumed model: No additional MCMC (or other numerical) simulations were required to obtain these results.

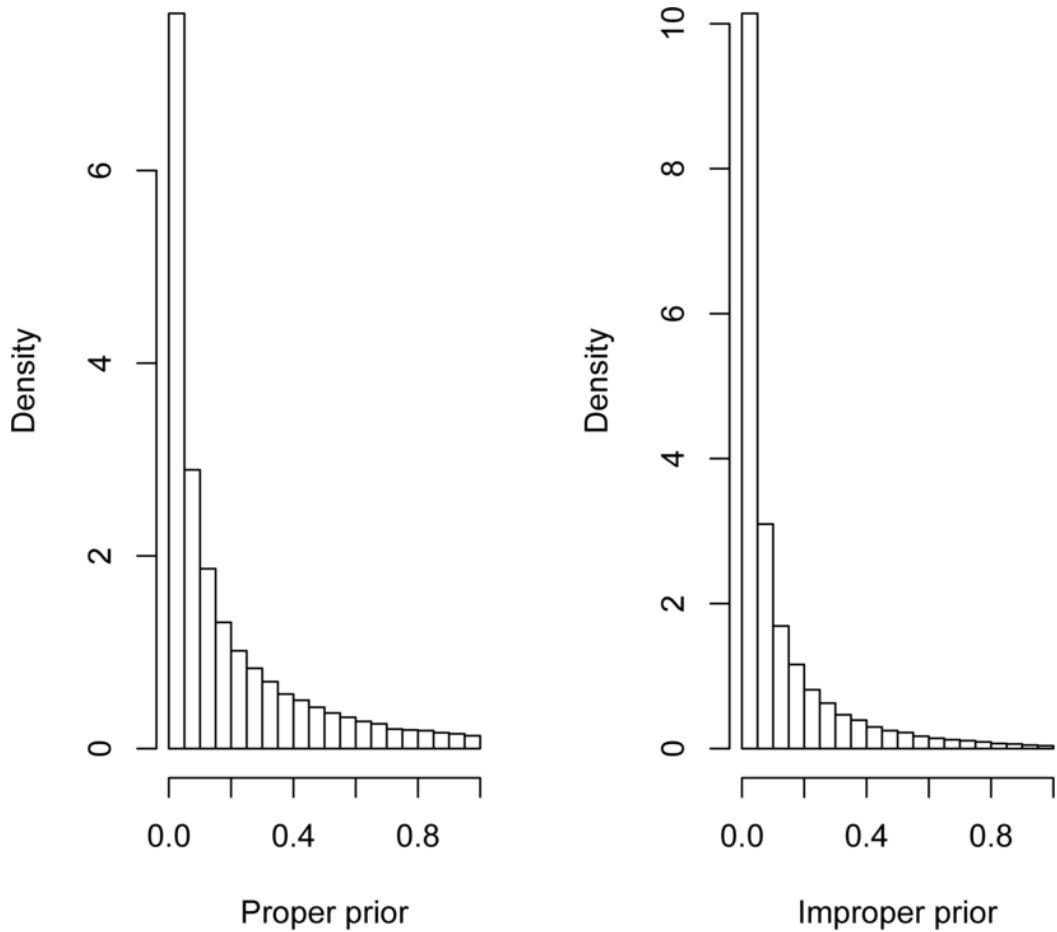

FIG. 2. *p values obtained by applying the Shapiro–Wilks test statistic to second-stage model residuals. p values displayed in the left panel were obtained from a model based on a proper prior distribution; the right panel displays p values obtained from a model specified with an improper prior distribution.*

Turning now to the hospital mortality data, suppose that the Jeffreys prior assumed for $(\alpha, \beta)$ by B&C is truncated to the interval $(a, 1/a) \times (a, 1/a)$ for a suitably small value of $a$. When evaluated at independent samples of $\{p_i\}$, $\alpha$ and $\beta$ drawn from the posterior, it follows that values of $\zeta_i$ defined by

(3) $\qquad \zeta_i = Beta(p_i; \alpha, \beta), \quad i = 1, \ldots, 12,$

are marginally distributed as i.i.d. uniform deviates under the assumed model. Thus, model adequacy can be evaluated by either examining vectors of these uniform values in probability plots, or by transforming their values to a scale appropriate for the model at hand. To this end, Figure 3 displays three randomly selected quantile-quantile plots of posterior samples of $\{p_i\}$ against quantiles from the corresponding $Beta(\alpha, \beta)$ distribution. Each of these plots suggests that the hospital mortality rates may not have been generated from a common beta distribution.

Bayarri and Castellanos' selection of the maximum proportion as a test statistic to conduct partial posterior model checks can be mimicked here by selecting the largest uniform deviate from each posterior sample of quantities in (3) as a summary test statistic. It follows that for a single vector $\zeta_i$ drawn from the posterior, the twelfth order statistic, $\zeta_{(12)}$, has distribution function $F(x) = x^{12}$. Figure 4 displays a quantile-quantile plot of 250,000 $\zeta_{12}$ values drawn from the posterior against the corresponding expected order statistics.

Bounds on the distribution of dependent order statistics can again be applied to values displayed in Figure 4 to obtain a bound on the $p$ value for model fit. For this test statistic, a bound of $p < 0.10$ is obtained. As before, calculation of this bound requires only output available from the MCMC algorithm used to sample from the posterior distribution. No additional simulation experiments or numerical approximations are needed.

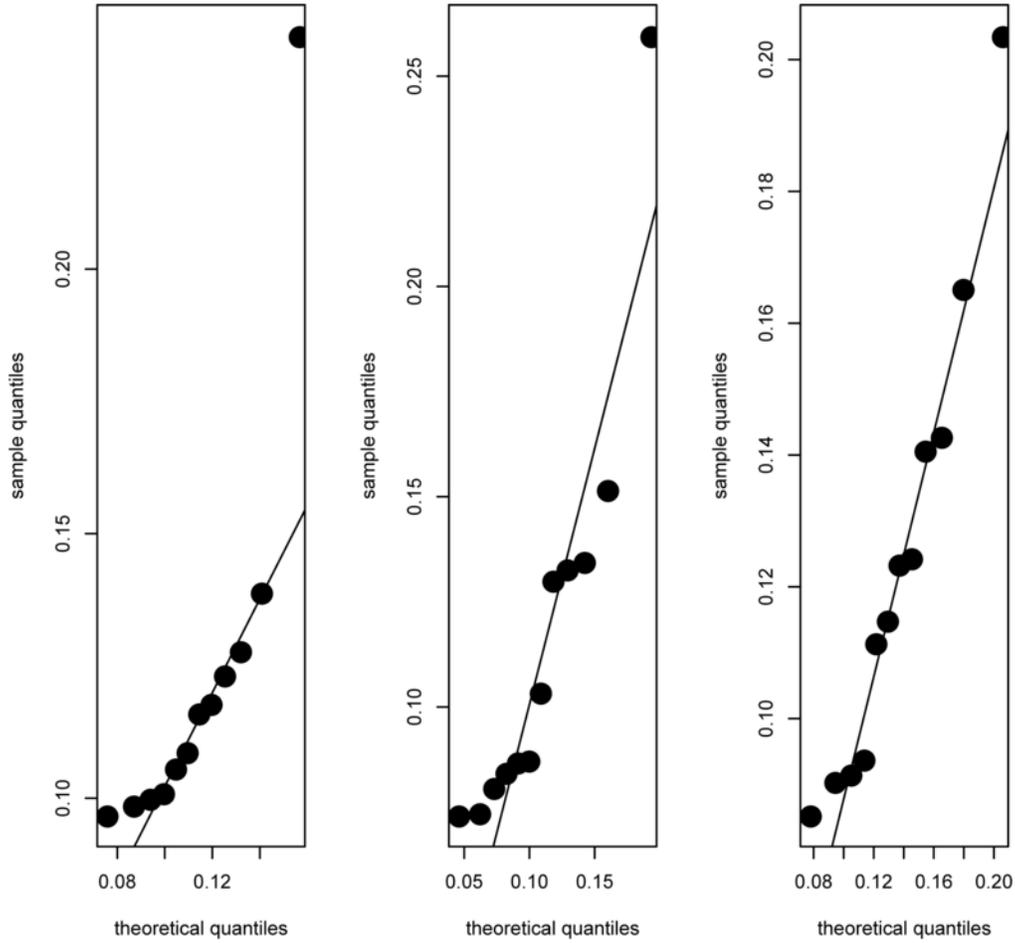

FIG. 3. *Quantile-quantile plots of hospitality mortality rates.*

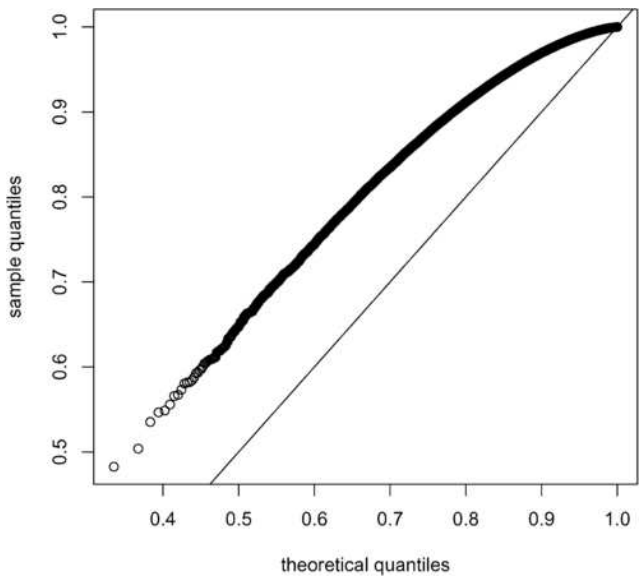

FIG. 4. *Quantile-quantile plot of largest uniform statistic $\zeta_{(12)}$ obtained from 250,000 posterior samples. The line indicated in the plot has slope 1 and intercept 0.*

Returning to a discussion of partial posterior $p$ values, methodologies proposed by B&C for assessing the adequacy of second levels of hierarchical models offer important advantages over several competing methods, but they also present several practical difficulties.

These difficulties include the following:

1. Numerical evaluation of partial posterior distributions is computationally and conceptually challenging even in simple normal theory problems. Defining appropriate test statistics and estimating partial posterior distributions in more complicated models may prove impracticable.

2. Nonuniformity of $p$ values in finite samples, coupled with the numerical approximation of partial posterior distribution of the observed test statistic, makes it difficult to assess the evidence con-tained in small $p$ values. As the authors point out, anticonservatism is probably better than conservatism when diagnosing model fit. But neither is good, and the relative error associated with small $p$ values is potentially quite large.

3. Propriety of partial posterior distributions may be difficult to establish when objective priors are employed, particularly when selected test statistics represent a component of a sufficient statistic.
4. Partial posterior model checks do not naturally facilitate graphical diagnostics and other informal model checks that are critical to the processes of model refinement and criticism.

Partial posterior $p$ values do, however, possess an important property not shared by many competing methods: Partial posterior $p$ values can substantially diminish the effects of masking. Indeed, evidence provided in the article suggests that partial posterior $p$ values are an order of magnitude less sensitive to masking than $p$ values computed using other standard methods. Provided that the proposed methodology can be extended to realistically complex Bayesian models, this property offers assurance that large deviations from model assumptions will not be overlooked simply because, say, a variance parameter was overestimated.

## REFERENCES

BAYARRI, M. J. and BERGER, J. O. (1999). Quantifying surprise in the data and model verification. In *Bayesian Statistics 6* (J. M. Bernardo, J. O. Berger, A. P. Dawid and A. F. M. Smith, eds.) 53–82. Oxford Univ. Press. MR1723493

CARAUX, G. and GASCUEL, O. (1992). Bounds on distribution function of order statistics for dependent variates. *Statist. Probab. Lett.* **14** 103–105. MR1173405

GELMAN, A., MENG, X. L. and STERN, H. S. (1996). Posterior predictive assessment of model fitness via realized discrepancies (with discussion). *Statist. Sinica* **6** 733–807. MR1422404

JOHNSON, V. E. (2007). Bayesian model assessment using pivotal quantities. *Bayesian Analysis* **2**. To appear.

O'HAGAN, A. (2003). HSSS model criticism (with discussion). In *Highly Structured Stochastic Systems* (P. J. Green, N. L. Hjort and S. T. Richardson, eds.) 423–445. Oxford Univ. Press. MR2082403

ROBINS, J. M., VAN DER VAART, A. and VENTURA, V. (2000). Asymptotic distribution of $p$ values in composite null models (with discussion). *J. Amer. Statist. Assoc.* **95** 1143–1172. MR1804240

RYCHLIK, T. (1992). Stochastically extremal distributions of order statistics for dependent samples. *Statist. Probab. Lett.* **13** 337–341. MR1175159

SHAPIRO, S. and WILKS, M. (1965). An analysis of variance test for normality: Complete samples. *Biometrika* **52** 591–611. MR0205384